\documentclass{ws-mpla}

\usepackage{braket}
\usepackage[abs]{overpic}
\usepackage{feynmf}

\newcommand{\comment}[1]{}

\newcommand{\babar}{\mbox{\slshape B\kern-0.1em{A}\kern-0.1em B\kern-0.1em{A\kern-0.2em R}}}
\newcommand{\belle}{Belle}

\newcommand{\etal}{\textit{et al.}}  

\newcommand{\CP}{\textit{CP}}                                            %	CP
\newcommand{\CC}{\textit{C}}                                       %	C
\newcommand{\PP}{\textit{P}}                                          %	P
\newcommand{\CPT}{\textit{CPT}}                                            %	CP

\newcommand{\Dhh}{\ensuremath{D^0 \to h^+ h^-}}

\newcommand{\im}{\ensuremath{{\rm \Im m}}}
\newcommand{\re}{\ensuremath{{\rm \Re e}}}
 \newcommand{\Acp}{\ensuremath{A_{\CP}}}

\begin{document}

\markboth{Michael J. Morello}
{Measurement of CP violation in $D^0/\bar{D}^0$}

%%%%%%%%%%%%%%%%%%%%% Publisher's Area please ignore %%%%%%%%%%%%%%
\catchline{}{}{}{}{}
%%%%%%%%%%%%%%%%%%%%%%%%%%%%%%%%%%%%%%%%%%%%%%%%%%%%%%%%%%%%%%%%%%%

\title{MEASUREMENT OF CP VIOLATION IN $D^0/\bar{D}^0$
%USING \TeX\ OR \LaTeX
%\footnote{For the title, try not to use more than 
%three lines. Typeset the title in 10 pt Times Roman, uppercase and 
%boldface.}
}

\author{\footnotesize MICHAEL J. MORELLO
%\footnote{
%Typeset names in 8 pt Times Roman, uppercase. Use the footnote to 
%indicate the present or permanent address of the author.}
}

\address{Scuola Normale Superiore, Piazza dei Cavalieri 7\\ 
Pisa, 56127, Italy\\
michael.morello@sns.it}
%University Department, University Name, Address\\
%City, State ZIP/Zone,
%Country\footnote{State completely without abbreviations, the 
%affiliation and mailing address, including country and e-mail address. 
%Typeset in 8 pt Times Italic.}\\
%author@emailaddress}

%\author{SECOND AUTHOR}

%\address{Group, Laboratory, Address\\
%City, State ZIP/Zone, Country
%}

\maketitle

\pub{Received (Day Month Year)}{Revised (Day Month Year)}

\begin{abstract}

Charm physics has played all along a central role in particle physics, however % as a probe for physics beyond the Standard Model.
the level of attention on it has tremendously  increased in the last years because of the observation 
of ``fast''  $D^0-\bar{D}^0$ flavour oscillations and because of very recent observed hints of  \CP\ violation. 
While in the past these would have been unambiguously interpreted as signs of New Physics, the revisitation of 
theoretical expectations, prompted by the latest experimental measurements,  makes the picture not clear.
This brief review covers the current status of \CP-violating measurements in the $D^0-\bar{D}^0$ system, 
both on the experimental and theoretical side. 

\end{abstract}

\ccode{PACS Nos.:3.20.Fc, 13.25.Ft, 14.40.Lb.}

\section{Introduction}

The \CP\ transformation combines the charge conjugation \CC\ with the parity \PP. 
Under \CC\ operator  particles and antiparticles are interchanged by conjugating
all internal quantum numbers ( e.g., $Q \to -Q$ for electromagnetic charge), while  under \PP\ the handedness
of space is reversed, $x \to -\bar{x}$. 
So far most phenomena observed in Nature are \CC- and \PP-symmetric, and therefore, also \CP-symmetric. 
Gravitational, electromagnetic, and strong interactions are invariant under  \CC, \PP\ and then under the  \CP\ transformation, 
while the weak interactions  violate \CC\ and \PP\ separately in the strongest possible way. 
For a long time physicists believed that  weak interactions were \CP-symmetric, since 
the invariance under the \CP\ operator  is preserved in most weak processes.
However the \CP\ symmetry is also violated in certain rare weak processes, as discovered in neutral $K$ decays in 1964\cite{cpv_kmeson}, and observed 
in recent years in $B$ decays\cite{Aubert:2001nu,Abe:2001xe}.
These effects are related to $K^0 - \bar{K}^0$ and $B^0 - \bar{B}^0$ mixing, 
but CP violation arising solely from decay amplitudes has also been observed, first in $K \to \pi\pi$ decays \cite{Burkhardt:1988yh,Fanti:1999nm,ktev} and
more recently in various neutral\cite{Aubert:2004qm,belle_acp} and charged\cite{Aubert:2005ce,Garmash:2005rv} $B$ decays. 

Within the Standard Model the \CP\ symmetry is broken through the well-known Kobayashi-Maskawa (KM) mechanism \cite{km_mechanism},
by a single complex phase which appears in the $3 \times 3$ unitary matrix that gives the $W$-boson couplings 
to an up-type antiquark and a down-type quark, in the basis of mass eigenstates. 
The theory agrees with all measurements to date,  %Furthermore, one can fit the data allowing new physics contributions to loop processes 
providing a strong proof that the Cabibbo-Kobayashi-Maskawa (CKM) phase is different from zero, and that the matrix of three-generation quark mixing 
is the dominant source of \CP\ violation observed in the meson decays. However this is not sufficient  to explain the matter-antimatter asymmetry observed in our Universe. That asymmetry tells us 
New Physics (NP) with \CP\ violation has to exist. %The question is where can we find manifestation of such NP and at which level.
The usual candidate for finding the dynamics underlying the Universe’s baryonic asymmetry is neutrino oscillation with \CP\ violation, 
however the  heavy flavour sectors have not yet been fully covered by experiments so far.
For instance, the Standard Model (SM) predicts very small \CP\
violation for charm decays, therefore 
the dynamics of this quark can be well probed for the existence and analyses of NP without too much SM ``background''. 
Moreover the neutral $D$ mesons system is the only  one where up-sector quarks are involved in the initial state. 
Thus it probes scenarios where up-type quarks play 
a special role, such as supersymmetric models where the down quark and the squark mass matrices are aligned\cite{Nir:1993mx,Ciuchini:2007cw} and, 
more generally, models in which CKM mixing is generated in the up-quark sector. The interest in charm dynamics has increased recently with the observation 
of charm oscillations\cite{Aubert:2007wf,Staric:2007dt,Aaltonen:2007uc}. 
The current measurements\cite{hfag} indicate $\mathcal{O}(10^{-2})$ magnitudes for the parameters 
governing their phenomenology. Such values are on the upper end of most theory predictions\cite{Petrov:2006nc}. Charm oscillations could be enhanced 
by a broad class of non-SM physics processes\cite{Golowich:2007ka}. Any generic non-SM contribution to the mixing would naturally carry additional 
\CP-violating phases, which could enhance the observed \CP\ violating asymmetries relative to SM predictions. 

\section{Formalism}
\label{notazione}

The decay amplitudes of a $D$ meson (charged or neutral) and its \CP\ conjugate $\overline{D}$ to a 
multi--particle final state $f$ and its \CP\ conjugate $\overline{f}$ are defined as
\begin{equation}
A_f = \bra{f}\mathcal{H}\ket{D},\quad \overline{A}_f = \bra{f}\mathcal{H}\ket{\overline{D}},\quad
A_{\overline{f}} = \bra{\overline{f}}\mathcal{H}\ket{D},\quad
\overline{A}_{\overline{f}} = \bra{\overline{f}}\mathcal{H}\ket{\overline{D}}
\end{equation}
where $\mathcal{H}$ is the decay Hamiltonian.
There are two types of phases that may appear in $A_f$ and $\overline{A}_{\overline{f}}$. Complex parameters in 
any Lagrangian term that contributes to the amplitude will appear in complex conjugate form in the 
\CP--conjugate amplitude. Thus their phases appear in $A_f$ and $\overline{A}_{\overline{f}}$ with opposite signs. 
In the Standard Model these phases occur only in the CKM matrix which is part of the electroweak sector of the theory, 
hence these are often called ``weak phases''. 
%The weak phase of any single term is convention dependent. 
%However the difference between the weak phases in two different terms in $A_f$ is convention independent because
%the phase rotations of the initial and final states are the same for every term. 
A second type of phase can appear in 
scattering or decay amplitudes even when the Lagrangian is real. Such phases do not violate \CP\ and they 
appear in $A_f$ and $\overline{A}_{\overline{f}}$ with the same sign. Their origin is the possible contribution from intermediate 
on--shell states in the decay process, that is an absorptive part of an amplitude that has contributions 
from coupled channels. Usually the dominant rescattering is due to strong interactions and hence the 
designation ``strong phases'' for the phase shifts so induced. 
%Again only the relative strong phases of 
%different terms in a scattering amplitude have physical content, an overall phase rotation of the entire amplitude has no physical consequences.

\CP\ violation in the decay appears as a result of interference among various terms in the decay amplitude, and will not occur unless 
at least two terms have different weak phases and different strong phases. As an example, let us consider a decay process which 
can proceed through several amplitudes:
$$A_f = \sum_j |A_j|\ e^{i(\delta_j+\phi_j)},\qquad
\overline{A}_{\overline{f}} = \sum_j |A_j|\ e ^{i(\delta_j-\phi_j)},$$
where $\delta_j$ and $\phi_j$ are strong (\CP\ conserving) and weak (\CP\ violating) phases, respectively. To observe \CP\ violation 
one needs $|A_f|\neq|\overline{A}_{\overline{f}}|$, i.e,~there must be a contribution from at least two processes with different 
weak and strong phases in order to have a non vanishing interference term
$$|A_f|^2 -|\overline{A}_{\overline{f}}|^2=-2\sum_{i,j}|A_i||A_j|\sin(\delta_i-\delta_j) \sin(\phi_i-\phi_j).$$

The phenomenology of \CP\ violation in neutral flavored meson decays is enriched by the possibility that, 
besides the decay, it is also possible to have  $D^0\leftrightarrow\overline{D}^0$ transitions, 
also known as flavor mixing or oscillations. Particle--antiparticle mixing has been observed in all 
four flavored neutral meson systems, i.e., in neutral kaon, both neutral $B$ meson systems and neutral $D$ meson system. 
The particle--antiparticle mixing phenomenon causes an initial (at time $t=0$), pure $D^0$ meson state to evolve in 
time to a linear combination of $D^0$ and $\overline{D}{}^0$ states. If the times $t$ in which we are interested are much 
larger than the typical strong interaction scale, then the time evolution can be described by the approximate Schr\"odinger equation:
\begin{equation}
i\frac{d}{dt}
\begin{pmatrix}
D^0(t)\\
\overline{D}^0(t)
\end{pmatrix}
=\left[\bm{M}-\frac{i}{2}\bm{\Gamma}\right]
\begin{pmatrix}
D^0(t)\\
\overline{D}^0(t)
\end{pmatrix},
\end{equation}
where $\mathbf{M}$ and $\mathbf{\Gamma}$ are $2\times 2$ Hermitian matrices,
$$\bm{M}=\begin{pmatrix}
M_{11} & M_{12}\\
M_{12}^\ast & M_{22}\\
\end{pmatrix}\quad\text{and}\quad
\bm{\Gamma}=\begin{pmatrix}
\Gamma_{11} & \Gamma_{12}\\
\Gamma_{12}^\ast & \Gamma_{22}\\
\end{pmatrix},$$
associated with transitions via off--shell (dispersive) and on--shell (absorptive) intermediate states, respectively. 
Diagonal elements of $\mathcal{H}_{\rm eff}=\bm{M}-i\bm{\Gamma}/2$ are associated with the flavor--conserving 
transitions $D^0\to D^0$ and $\overline{D}^0\to\overline{D}^0$ while off--diagonal elements are associated with 
flavor--changing transitions $D^0\leftrightarrow\overline{D}^0$. The matrix elements of $\bm{M}$ and $\bm{\Gamma}$ 
must satisfy $M_{11}=M_{22}$ and $\Gamma_{11}=\Gamma_{22}$ in order to be consistent with \CPT\ invariance.

The eigenstates of the effective Hamiltonian $\mathcal{H}_{\rm eff}$ are
$$\ket{D_{L,H}} = p\ \ket{D^0}\pm q\ \ket{\overline{D}^0}$$
while the corresponding eigenvalues are
$$\lambda_{L,H}=\left (M_{11}-\frac{i}{2}\Gamma_{11}\right)\pm\frac{q}{p}\left({M}_{12}-\frac{i}{2}{\Gamma}_{12}\right)\equiv m_{L,H}-\frac{i}{2}\Gamma_{L,H}.$$
The coefficients $p$ and $q$ are complex coefficients, satisfying $|p|^2+|q|^2=1$, and
$$\frac{q}{p}=\sqrt{\frac{M_{12}^{\ast}-\frac{i}{2}\Gamma_{12}^{\ast}}{M_{12}-\frac{i}{2}\Gamma_{12}}}=\left|\frac{q}{p}\right|e^{i\phi}.$$
The real parts of the eigenvalues $\lambda_{1,2}$ represent masses, $m_{L,H}$, and their imaginary 
parts represent the widths $\Gamma_{L,H}$ of the two eigenstates $\ket{D_{L,H}}$, respectively. 
The sub--scripts $H$ (heavy) and $L$ (light) are here used because by convention 
we choose $\Delta m= m_H-m_L>0$, while the sign of $\Delta\Gamma = \Gamma_L-\Gamma_H$ is not 
known \textit{a priori} and needs to be experimentally determined. 
%\footnote{Another possible choice, which is in standard 
%usage for $K^0$ mesons, defines the mass eigenstates according to their lifetimes: $K_\text{S}$ for the short--lived 
%and $K_\text{L}$ for the long--lived state, with $\Delta\Gamma_K=\Gamma_\text{S}-\Gamma_\text{L}$ positive by definition. 
%The $K_\text{L}$ is then experimentally found to be the heavier state, i.e.,~also $\Delta m_K>0$.}

The time--dependent decay amplitude of an initially pure $D^0$ state decaying to final state $f$ is then given by
$$\bra{f}\mathcal{H}\ket{D^0(t)}=A_f\ g_+(t)+\bar{A}_f\ \frac{q}{p}\ g_-(t),$$
where
$$|g_\pm(t)|^2=\frac{1}{2}\ e^{-t/\tau}\left[\cos\left(\frac{xt}{\tau}\right)\pm\cosh\left(\frac{yt}{\tau}\right)\right]$$
represents the time--dependent probability to conserve the initial flavor ($+$) or oscillate into the opposite flavor ($-$) and $x$, $y$ are 
dimensionless mixing parameters defined as
%\begin{align*}
% x &= \frac{\Delta m}{\Gamma},\\
% y &= \frac{\Delta\Gamma}{2\Gamma},
%\end{align*}
$$x = \frac{\Delta m}{\Gamma},\qquad y = \frac{\Delta\Gamma}{2\Gamma},$$
and $\Gamma = (\Gamma_L+\Gamma_H)/2=1/\tau$ is the mean decay width.

The time--dependent decay rate, proportional to $|\bra{f}\mathcal H \ket{D^0(t)}|^2$, is then
\begin{multline*}
\frac{d\Gamma}{dt}(D^0(t)\to f) \propto |A_f|^2\bigg[(1-|\lambda_f|^2)\cos\left(\frac{xt}{\tau}\right)+(1+|\lambda_f|^2)\cosh\left(\frac{yt}{\tau}\right)\\
-2\im(\lambda_f)\sin\left(\frac{xt}{\tau}\right)+2\re(\lambda_f)\sinh\left(\frac{yt}{\tau}\right)\bigg].
\end{multline*}
with
$$\lambda_f = \frac{q}{p}\ \frac{\overline{A}_f}{A_f}.$$
In analogy with this treatment one can show that for an initial pure $\overline{D}^0$ eigenstate the decay rate is
\begin{multline*}
\frac{d\Gamma}{dt}(\overline{D}^0(t)\to f) \propto |\bar{A}_f|^2\bigg[(1-|\lambda_f^{-1}|^2)\cos\left(\frac{xt}{\tau}\right)
+(1+|\lambda_f^{-1}|^2)\cosh\left(\frac{yt}{\tau}\right)\\
-2\im(\lambda_f^{-1})\sin\left(\frac{xt}{\tau}\right)+2\re(\lambda_f^{-1})\sinh\left(\frac{yt}{\tau}\right)\bigg].
\end{multline*}
Decay rates to the \CP--conjugate final state $\bar{f}$ are obtained analogously, with the substitutions $A_f\to A_{\overline{f}}$ 
and $\overline{A}_f\to\overline{A}_{\overline{f}}$ in the above equations. Terms proportional to $|A_f|^2$ or $|\overline{A}_f|^2$ are
 associated with decays that occur without any net $D^0\leftrightarrow\overline{D}^0$ oscillation, while terms proportional 
to $|\lambda_f|^2$ or $|\lambda_f^{-1}|^2$ are associated with decays following a net oscillation; the $\sin(xt/\tau)$ 
and $\sinh(yt/\tau)$ terms are instead associated with the interference between these two cases.

While \CP\ violation in charged meson decays depends only on $A_f$ and $\overline{A}_{\overline{f}}$, in the case of neutral mesons, 
because of the possibility of flavor oscillations, \CP\ violating effects have additional dependences 
on the values of $|q/p|$ and $\lambda_f$. We then distinguish three types of \CP\ violating effects in meson decays: 
\begin{enumerate}[\itshape (i)]
\item \CP\ violation in the decay is defined by
$$|\overline{A}_{\overline{f}}/A_f|\neq1.$$
In charged meson decays, where mixing effects are absent, this is the only possible source of CP asymmetries:
\begin{equation}\label{acp-direct}
\Acp(D\to f)\equiv\frac{\Gamma(D\to f)-\Gamma(\overline{D} \to \overline{f})}{\Gamma(D\to f)+\Gamma(\overline{D}\to\overline{f})}
=\frac{1-|\overline{A}_{\overline{f}}/A_f|^2}{1+|\overline{A}_{\overline{f}}/A_f|^2}
\end{equation}
\item \CP\ violation in mixing is defined by
$$|q/p|\neq1.$$
In this case, in place of Eq.~\ref{acp-direct}, it is useful to define the time--dependent asymmetry
\begin{equation}\label{acp-time-dependent}
\Acp(D^0\to f; t) =\frac{d\Gamma(D^0(t)\to f)/dt-d\Gamma(\overline{D}^0(t) \to\overline{f})/dt}{d\Gamma(D^0(t)\to f)/dt
+d\Gamma(\overline{D}^0(t) \to\overline{f})/dt},
\end{equation}
\item CP violation in interference between a decay without mixing, $D^0\to f$, and a decay with mixing, $D^0\to\overline{D}^0\to f$ (such an effect occurs 
only in decays to final states that are common to both $D^0$ and $\overline{D}^0$, including all \CP\ eigenstates), is defined by
$$\im{\lambda_f}\ne 0$$
\end{enumerate}
Usually type (1) is also know as direct \CP\ violation, while type (2) and (3) are referred as indirect \CP\ violation.

\section{Neutral charmed mesons decays: D mixing \label{cpv-charm}}

\begin{figure}[t]
\centering
\vskip0.5cm
%\begin{overpic}[angle=180,width=0.95\textwidth,grid=false]{./fig/mix-sheldon-crop.pdf}
\begin{overpic}[angle=180,width=0.95\textwidth,grid=false]{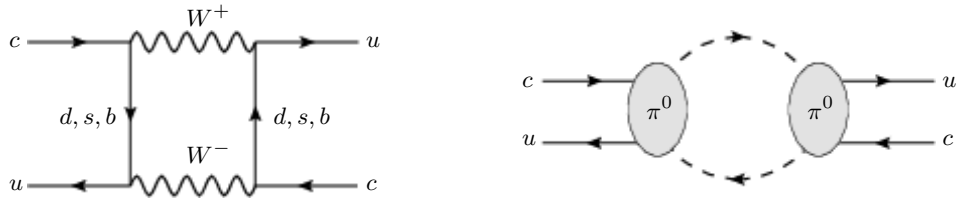}
%\begin{overpic}[angle=180,width=0.2\textwidth,grid=false]{mplaf1.eps}
%left
\put(-7,57){\small $c$}
\put(-7,3){\small $u$}
\put(128,57){\small $u$}
\put(128,3){\small $c$}
\put(12,28){\small $d,s,b$}
\put(93,28){\small $d,s,b$}
\put(60,65){\small $W^+$}
\put(60,13){\small $W^-$}
%right
\put(187,42){\small $c$}
\put(187,20){\small $u$}
\put(346,42){\small $u$}
\put(346,20){\small $c$}
\put(234,30){\small $\pi^0$}
\put(295,30){\small $\pi^0$}
\end{overpic}
\vskip0.4cm
\caption{\em Examples of Feynman diagrams which describe ``short'' (left) and ``long distance'' (right) contributions to the $D^0-\overline{D}^0$ 
mixing amplitude. In the Standard Model the latter diagrams dominate over the ``short distance'' ones which are 
negligible compared to the first because of the small CKM coupling to the $b$ quark and of GIM suppression of the remaining two light--quark loops.}
\label{dmix}
\end{figure}

The interest in charm dynamics has increased recently with the evidence of charm oscillations reported by 
three different experiments~\cite{Aubert:2007wf,Staric:2007dt,Aaltonen:2007uc}, which, when combined together with 
all other available experimental information, established the existence of mixing at the $10\sigma$ level~\cite{hfag}. 
In the Standard Model mixing in neutral $D$ meson system can proceed through a double weak boson exchange 
(short distance contributions) represented by box diagrams, or through intermediate states that are accessible to both $D^0$ and 
$\overline{D}^0$ (long distance effects), as represented in Fig.~\ref{dmix}. Potentially large long distance contributions are 
non--perturbative and therefore difficult to estimate, hence the predictions for the mixing parameters $x$ and $y$ within the Standard Model 
span several orders of magnitude between $10^{-8}$ and $10^{-2}$ \cite{Petrov:2006nc}. The measured values of $x$ and $y$, 
as averaged by the \textit{Heavy Flavor Averaging Group} (HFAG) when CP violation is allowed \cite{hfag}, are
\begin{equation}\label{charm-mixing}
x=(0.63^{+0.19}_{-0.20})\%\quad\text{and}\quad y=(0.75\pm0.12)\%.
\end{equation}
The large uncertainties of the Standard Model mixing predictions make it difficult to identify New Physics 
contributions (a clear hint would be, if $x$ is found to be much larger than $y$), however since current measurements 
are on the upper end of most theory predictions \cite{Petrov:2006nc}, they could be interpreted as a possible hint for New Physics.

Charm oscillations could be enhanced by a broad class of non--Standard Model physics processes \cite{Golowich:2007ka}: 
\i.e.,~models with extra fermions like a forth generation down--type quark, with flavor changing neutral 
currents at tree level mediated by additional gauge bosons or in general with new symmetry of the 
theory like in Supersymmetry (SUSY). Any generic New Physics contribution to the mixing would naturally 
carry additional CP--violating phases, which could enhance the observed CP--violating asymmetries relative 
to Standard Model predictions. Moreover, since charmed hadrons are the only hadrons, presently accessible 
to experiment, composed of a heavy charged $+2/3$ quark\footnote{The top quark decays before it forms a hadron 
and therefore cannot oscillate; the absence of flavor mixing reduces significantly the possibility to study \CP\ violating effects 
involving the other down--type quarks.}, they provides the sole window of opportunity to examine scenarios where up--type quarks 
play a special role, such as SUSY models where the down quark and the squark mass matrices are aligned~\cite{Nir:1993mx,Ciuchini:2007cw} and, 
more generally, models in which CKM mixing is generated in the up--quark sector.

\section{Cabibbo--suppressed $D^0\to\pi^+\pi^-$ and $D^0\to K^+K^-$ decays\label{scs-theory}} \label{dhh_teoria}
Examples of clean channels where to study both direct and indirect CP violation in the charm system are the neutral 
singly--Cabibbo--suppressed decays into CP--eigenstates, such as $D^0\to\pi^+\pi^-$ and $D^0\to K^+ K^-$ 
(collectively referred to as \Dhh\ in the following).
Owing to the slow mixing rate of charm mesons, the time--dependent asymmetry of Eq.~\ref{acp-time-dependent} 
can be approximated to first order as the sum of two terms:
\begin{equation}\label{acp-time-dependent-charm}
\Acp(D^0\to f;t) = \Acp^{\rm dir}(D^0\to f)+\frac{t}{\tau}\ \Acp^{\rm ind}(D^0\to f)\quad(x,y\ll\tau/t)
\end{equation}
where $\Acp^{\rm dir}$ and $\Acp^{\rm ind}$ represents direct and indirect CP asymmetries, respectively. In the case $f$ is a CP eigenstate, 
as for \Dhh\ decays, then
\begin{align}
\Acp^{\rm dir}(D^0\to f)  &=\frac{1-\left|\overline{A}_{f}/A_f\right|^2}{1+\left|\overline{A}_{f}/A_f\right|^2}\quad\text{and}\\
\Acp^{\rm ind}(D^0\to f) &=\frac{1}{2}\left[y\ \re(\lambda_f-\lambda_f^{-1}) - x\ \im(\lambda_f-\lambda_f^{-1})\right]
\end{align}

\begin{figure}[t]
\centering
\vskip0.4cm
%\begin{overpic}[width=0.9\textwidth,grid=false]{./fig/graph-crop.pdf}
\begin{overpic}[width=0.9\textwidth,grid=false]{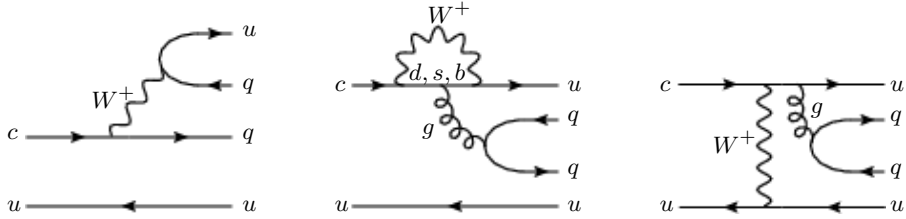}
%\begin{overpic}[width=0.2\textwidth,grid=false]{mplaf1.eps}
%left
\put(-7,28){\small$c$}
\put(-7,1){\small$u$}
\put(82,1){\small$u$}
\put(82,28){\small$q$}
\put(82,47){\small$q$}
\put(82,67){\small$u$}
\put(25,40){\small $W^+$}
%center
\put(117,46.5){\small$c$}
\put(117,1){\small$u$}
\put(205,1){\small$u$}
\put(205,46.5){\small$u$}
\put(205,15){\small$q$}
\put(205,35){\small$q$}
\put(152,72){\small $W^+$}
\put(145,50.5){\small $d,s,b$}
\put(150,30){\small $g$}
%right
\put(240,46.5){\small$c$}
\put(240,1){\small$u$}
\put(327,1){\small$u$}
\put(327,46.5){\small$u$}
\put(327,15){\small$q$}
\put(327,35){\small$q$}
\put(260,23){\small $W^+$}
\put(297,38){\small $g$}
\end{overpic}
\vskip0.4cm
\caption{\em Feynman diagrams of the possible topologies (from left to right: tree, one--loop penguin, 
$W$--exchange) of the \Dhh\ decays. The symbol $q$ stands for either a $d$ or a $s$ quark.}\label{graphs}
\end{figure}

Within the Standard Model direct CP violation can occur in singly--Cabibbo--suppressed charm 
decays ($c\to uq\overline{q}$ with $q=d,s$) because the final state particles contain at least one pair 
of quark and antiquark of the same flavor, which makes a contribution from penguin--type or box 
amplitudes induced by virtual $b$--quarks possible in addition to the tree 
amplitudes~\cite{Bianco:2003vb}.\footnote{Conversely, in the Standard Model, it is not possible to have direct 
CP violation in Cabibbo--favored ($c\to su\overline{d}$) or doubly--Cabibbo--suppressed ($c\to du\overline{s}$) charm decays.} 
However, as shown in the Feynman diagrams of Fig.~\ref{graphs}, the contribution of these second order amplitudes are 
strongly suppressed by the small combination of CKM matrix elements $V_{cb}V_{ub}^*$. Moreover, the tree amplitudes 
are practically CP conserving, since, for both $D^0\to\pi^+\pi^-$ and $D^0\to K^+K^-$ decays, they involve only one 
CKM factor, $V_{cd} V^*_{ud}$ and $V_{cs} V^*_{us}$ respectively, which is real in Wolfenstein parametrization 
up to $\mathcal{O}(\lambda^4)$ and $\mathcal{O}(\lambda^6)$. Hence to first order one would expect to observe an asymmetry 
consistent only with the mixing phase $\phi$, with no decay phase contribution:
%\begin{align*}
%\Acp(\Dhh) &\approx\Acp^{\rm ind}(\Dhh)\\
%&\approx \frac{\eta_{\rm CP}}{2}\left[y \left(\left|\frac{q}{p}\right|-\left|\frac{p}{q}\right|\right)\cos\phi-x \left(\left|\frac{q}{p}\right|+\left|\frac{p}{q}\right|\right)\sin\phi\right],
%\end{align*}
\begin{align*}
\Acp(\Dhh) &\approx\Acp^{\rm ind}(\Dhh)\\
&\approx \frac{\eta_{\rm CP}}{2}\left[-y \left(\left|\frac{q}{p}\right|-\left|\frac{p}{q}\right|\right)\cos\phi+x \left(\left|\frac{q}{p}\right|+\left|\frac{p}{q}\right|\right)\sin\phi\right],
\end{align*}
where $\eta_{\rm CP}=+1$ is the CP eigenvalue of the $h^+h^-$ final state. The Standard Model dynamics predicts indirect 
CP asymmetries around $\mathcal{O}(10^{-3})$, being suppressed by the value of $x$ and $y$ (see Eq.~\ref{charm-mixing}), while direct CP violation 
produces asymmetries one order of magnitude smaller. In addition, in the limit of $U$--spin symmetry, the direct component is equal 
in magnitude and opposite in sign for $D^0\to K^+K^-$ and $D^0\to\pi^+\pi^-$ \cite{Grossman:2006jg}.

As already mentioned in the previous section, New Physics contributions to the charm mixing would, in general, also exhibit larger CP violation. 
This mixing--induced effects, in many scenarios beyond the Standard Model, would in addition provide sources of direct CP violation 
in \Dhh\ decays both at tree level (extra quark in Standard Model vector--like representation, SUSY without R--parity models, 
two Higgs doublet models) or at one--loop (QCD penguin and dipole operators, flavor changing neutral currents in 
supersymmetric flavor models) as described in Ref.~\cite{Grossman:2006jg}. While the first group of models can produce 
an effect that is much less than 1\%, the processes having one--loop can even reach the percent level, producing 
effects that are clearly not expected in the Standard Model.

In the absence of large new weak phases in the decay amplitudes, i.e.,~negligible direct CP violation from New Physics, 
the CP asymmetries in singly--Cabibbo--suppressed decays into final CP eigenstates would be dominated by mixing--induced effects 
and thus universal, i.e.,~independent of the final state. So if different asymmetries are observed between $D^0\to\pi^+\pi^-$ and $D^0\to K^+K^-$ decays,  
then direct CP violation must be present.

\section{Measurement of time--integrated CP asymmetries}
The sources of a possible asymmetry in neutral $D$ meson decays can be distinguished by their dependence on the 
decay--time, so a time--dependent analysis seems necessary. However sensitivity to indirect CP violation 
can be achieved also with time--integrated measurements, if the detector acceptance allows to collect samples of $D^0$ mesons with 
decay--times longer than $\tau$. In fact, the time--integrated asymmetry is the integral of Eq.~\ref{acp-time-dependent-charm} 
over the observed distribution of proper decay time, $D(t)$:
\begin{multline}\label{acp-time-integrated}
\Acp(\Dhh) = \Acp^{\rm dir}(\Dhh)+\Acp^{\rm ind}(\Dhh)\int_0^\infty \frac{t}{\tau}\ D(t)dt\\
=\Acp^{\rm dir}(\Dhh) + \frac{\langle t \rangle}{\tau}\ \Acp^{\rm ind}(\Dhh).
\end{multline}
Since the value of $\langle t \rangle$ depends on $D(t)$, different values of time--integrated asymmetry may be observed in 
different experimental environments because of different detector acceptances as a function of decay time, thus providing different 
sensitivities to $\Acp^{\rm dir}$ or $\Acp^{\rm ind}$. In experiments where the reconstruction efficiency does not depend on proper decay 
time ($D(t)=1$), as it is the case at the $B$--factories, the factor $\langle t \rangle /\tau$ equals unity resulting in the same sensitivity to 
direct and indirect CP violation. On the contrary, %as we will detail in Sec.~\ref{trigger},
in experiments where data are collected with an online event selection (trigger), that imposes requirements on the displacement 
of the $D^0$ meson decay point from the production point, %thus rejecting candidates with short decay times, 
as it is the case at the CDF and LHCb, 
results $\langle t  \rangle /\tau>1$. This makes measurements, performed in hadronic environment, as CDF and LHCb,  more sensitive 
to mixing--induced CP violation. In addition, combination of these results with those from \belle\ and \babar\ 
provides discrimination between the two contributions to the asymmetry.

From the experimental point of view the flavour of neutral $D$ mesons at production is tagged by
reconstructing $D^{\ast +}\to D^0\pi_s^+$ decays
%\footnote{Charge conjugated processes are implied throughout the paper, unless  explicitly noted otherwise.} 
in which the charge of the low momentum
pion, $\pi_s$, determines the flavour of the $D^0$ meson. 
The measured asymmetry, 
$$A_{\rm rec}^{hh}=\frac{N(D^0\to h^+h^-)-N(\bar{D}^0\to h^+h^-)}{N(D^0\to h^+h^-)+N(\bar{D}^0\to h^+h^-)},$$ 
with $N$ denoting the number of reconstructed decays,
can be written, in general,  as a sum of several (assumed small) contributions:
\begin{equation}
%  A=A_{FB}+A_{CP}^f+A_\epsilon^\pi~.
  A^{hh}_{\rm rec}=A_{p}+A_{CP}^{hh}+A_\epsilon^\pi~.
  \label{eq2}
\end{equation}
where $A^{hh}_{CP}$ is the intrinsic CP asymmetry,   $A_\epsilon^\pi$  is a contribution due
to an asymmetry in the reconstruction efficiencies of oppositely
charged $\pi_s$  and $A_p$ a contribution from a production asymmetry depending on the experimental environment. 
Since the final state $h^+h^-$ is self-conjugate, its
reconstruction efficiency does not affect measured asymmetry $A_{\rm rec}^{hh}$. 

At the Tevatron,  charm and anticharm mesons are 
expected to be created in almost equal numbers. Since the overwhelming majority of them are produced by \CP--conserving 
strong interactions, and the $p\bar{p}$ initial state is \CP\ symmetric, any small difference between the abundance of 
charm and anti-charm flavor is constrained to be antisymmetric in pseudorapidity.  As a consequence, the
net effect of any possible charge asymmetry in the production cancels out ($A_p=0$), as long as the distribution of the decays
  is symmetric in pseudorapidity \cite{Aaltonen:2011se}.\\
In the production of $D^{\ast +}$ mesons in $e^+e^-\to c\bar{c}$, instead,  there is a forward-backward asymmetry, 
which arises from $\gamma-Z^0$ interference and higher order QED effects~\cite{afb,afb1,afb2}.   This term is an odd function of 
the cosine of the $D^{\ast +}$ production polar angle in the center-of-mass (CM) system ($\cos\theta^\ast$).
%\footnote{Symbols with an asterix   in the paper denote quantities
%  in the CM frame, while those without asterix denote quantities in the
%  laboratory frame.}  ($\cos\theta^\ast$). 
Since detector acceptance, in $e^+e^-$ machines,  is not symmetric with respect to $\cos\theta^\ast$,
the measurement is performed in bins of $\cos\theta^\ast$, allowing to correct for the acceptance and extract
both $A_{p}$ and $A_{CP}^{hh}$~\cite{Aubert:2007if,Staric:2008rx}.
 
One of the main experimental difficulty of these measurements comes from the small differences in the detection efficiencies of tracks 
of opposite charge $A_\epsilon^\pi$ which may lead, if not properly taken into account, to spuriously-measured charge asymmetries. 
Relevant instrumental effects include differences in interaction cross sections with matter between positive and negative 
low-momentum hadrons and the geometry of the main tracking system. This must 
be corrected to better than one per mil to match the expected statistical precision of the current measurements.
To reliably determine $A_\epsilon^\pi$ all experiments adopt a similar fully data-driven technique, based on an appropriate combination
 of charge-asymmetries observed in different event samples. In addition to the $D^0\to h^+h^-$ modes mentioned above, 
two $D^0\to K^-\pi^+$ samples are reconstructed: one consisting of $D$ mesons with tagged initial flavour, and one consisting
of untagged candidates. The measured asymmetries for these modes can be written as
\begin{eqnarray}
  \nonumber
  A_{\rm rec}^{\rm tag}~~~&=&A_{p}+A_{CP}^{K\pi}+A_\epsilon^{K\pi}+A_\epsilon^\pi~,\\
  A_{\rm rec}^{\rm untag}&=&A_{p}+A_{CP}^{K\pi}+A_\epsilon^{K\pi}~.
  \label{eq3}
\end{eqnarray}
A notable difference with (\ref{eq2}) is that this final state is
not self-conjugate and thus an additional term $A_\epsilon^{K\pi}$
appears as a consequence of a possible asymmetry
in the reconstruction efficiency of the $D^0 \to K^{-}\pi^{+}$ decays. %We first use 
The two measurements in (\ref{eq3}) are used to determine $A_\epsilon^\pi$ at the TeVatron 
where $A_{p}$ is null, and $A_\epsilon^\pi + A_{p}$ at the B-Factories.
The fact that $A_{p}$ is antisymmetric with respect to $\cos\theta^\ast$ and $A_{CP}^{hh}$ is
independent of this variable allows to disentangle the two contributions at the $e^+e^-$ environment. 
Then the result is inserted into (\ref{eq2}) to extract $A_{CP}^{hh}$.
%%%%%%%%%%%%%%%%%%%%%%%%%%%%%%%%%%%%%%%%%%%%%%%%%%%%%%%%%%%%%%%%%%%%%%%%%%%%%%%%%%
\begin{table}[h]
\tbl{Summary of recent experimental measurements of \CP\ violating asymmetries 
in two--body singly--Cabibbo--suppressed decays of $D^0$ mesons.} %\cite{Aubert:2007if,:2008rx,Acosta:2004ts}.} 
{\begin{tabular}{lcc}
\toprule 
Experiment      & $\Acp(D^0\to \pi^+\pi^-)\ (\%)$ & $\Acp(D^0\to K^+K^-)\ (\%)$ \\
\colrule 
\babar\ 2008 \cite{Aubert:2007if}      & $-0.24\pm0.52 \pm0.22  $ & $+0.00\pm0.34\pm0.13$ \\
\belle\ 2008 \cite{Staric:2008rx}        & $+0.43\pm0.52 \pm0.12 $ & $-0.43\pm0.30\pm0.11$ \\
CDF 2012 \cite{Aaltonen:2011se}       & $+0.22\pm0.24 \pm0.11 $ & $-0.24\pm0.22 \pm0.09$\\
\belle\ 2012 \cite{d0hh_belle_2012}   & $+0.55\pm0.36\pm0.09 $ & $-0.32\pm0.21\pm0.09$ \\
\botrule 
\end{tabular} \label{today}}
\end{table}
%%%%%%%%%%%%%%%%%%%%%%%%%%%%%%%%%%%%%%%%%%%%%%%%%%%%%%%%%%%%%%%%%%%%%%%%%%%%%%%%%%%%
% 
%%%%
%
\begin{figure*}[t]
\centering
\begin{overpic}[width=6.3cm,grid=false]{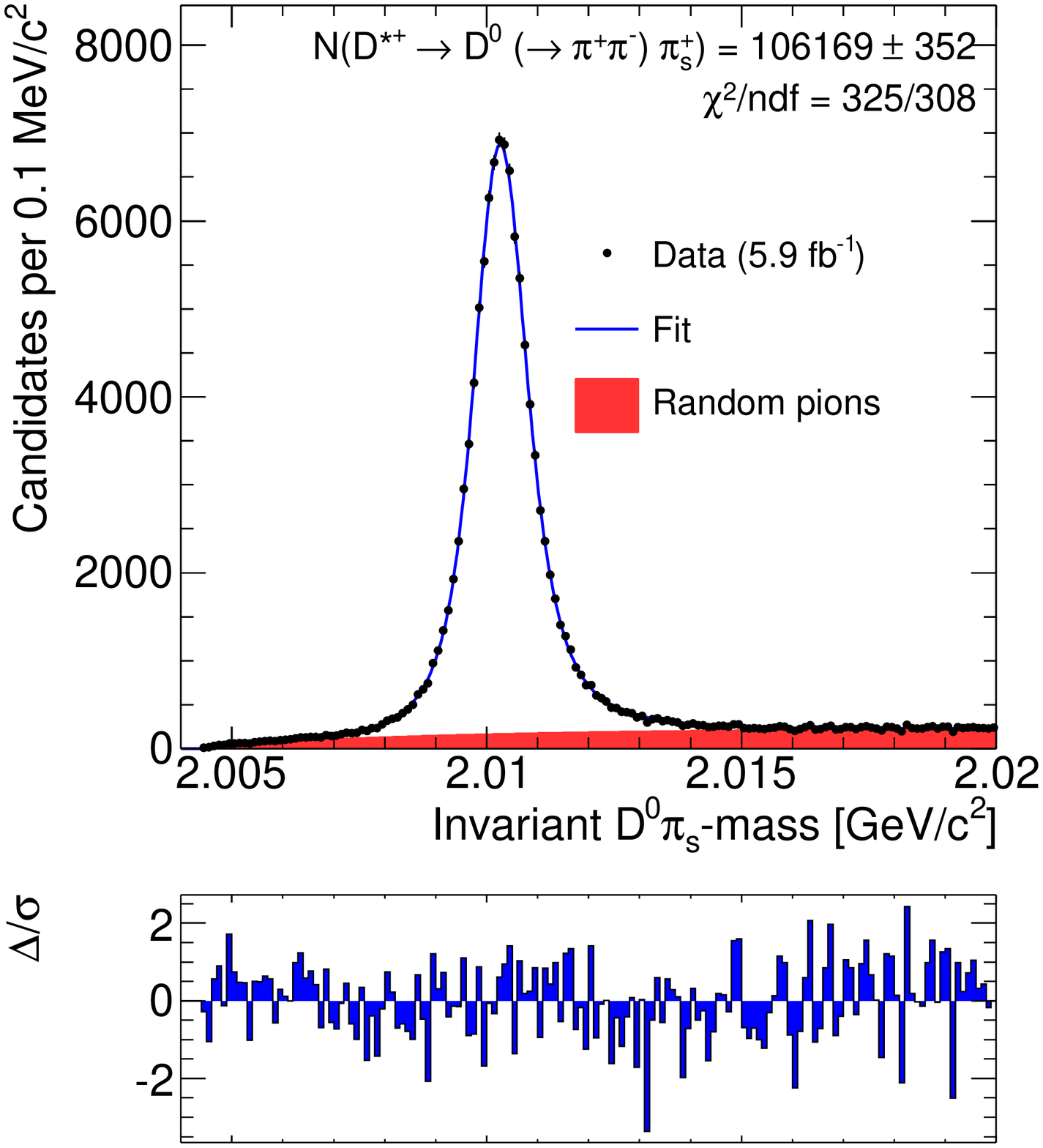}
\put(50,140){(a)} 
\end{overpic} \hfil
\begin{overpic}[width=6.3cm,grid=false]{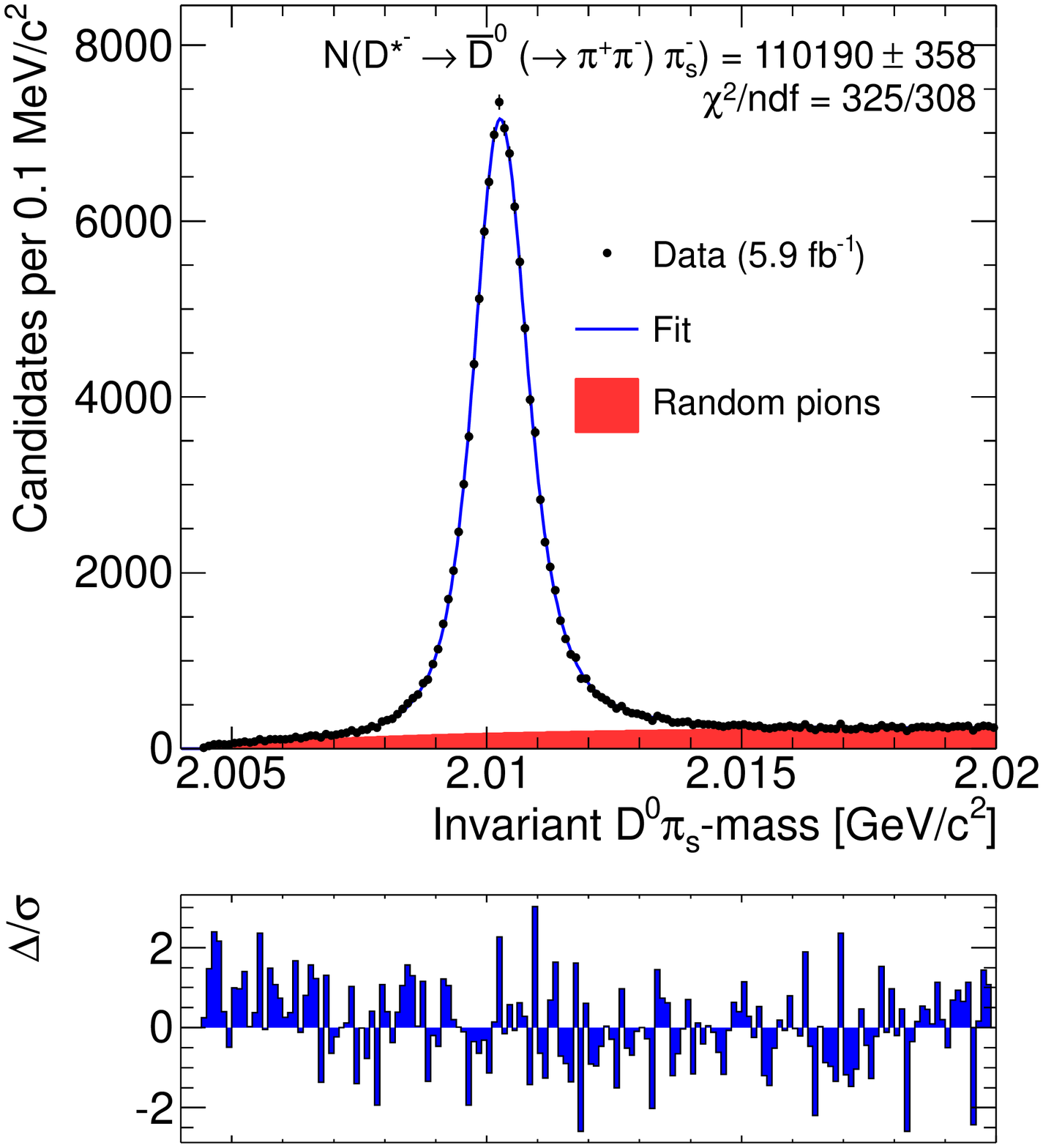}
\put(50,140){(b)}
\end{overpic}
%\caption[xxx]{dsdlsjlskjgld}
\caption[xxx]{Results of the combined fit of the tagged $D^0\to \pi^+\pi^-$ and  $\overline{D}^0\to \pi^+\pi^-$ samples at CDF~\cite{Aaltonen:2011se}. 
Distribution of $D^{0}\pi_s$ mass for (a) charm and (b)  anti-charm decays. Fit results are overlaid.}\label{fits-hh_cdf}
\end{figure*}
%%%%
The individual measurements of the time-integrated \CP\ asymmetries in the 
singly-Cabibbo-suppressed  decays into \CP-eigenstates are reported in Tab.~\ref{today}. 
%The world's average is dominated by recent CDF results~\cite{Aaltonen:2011se}. 
As reference we report the results of the combined fit of the tagged $D^0\to \pi^+\pi^-$ and  $\overline{D}^0\to \pi^+\pi^-$ samples at CDF in 
Fig.~\ref{fits-hh_cdf}, where the fit results are overlaid to the distribution of $D^{0}\pi_s$ mass. The signal yields 
are about 106 000 decays of $D^0\to \pi^+\pi^-$ and 110 000 of $\overline{D}^0\to \pi^+\pi^-$.

A useful comparison with results from different experiments is achieved by expressing the observed asymmetry 
as a linear combination  (Eq.\  (\ref{acp-time-integrated})) of a direct component,  $\Acp^{\rm{dir}}$,  and an 
indirect component, $\Acp^{\rm{ind}}$,  through a coefficient that is the mean proper decay time of 
charm mesons in the data sample used.  
%The direct component corresponds to a difference in width between 
%charm and anti-charm decays into the same final state. The indirect component is due to the probability for a 
%charm meson to oscillate into an anti-charm meson being different from the probability for an anti-charm meson to 
%oscillate into a charm meson.
%
\begin{figure*}[t]
\centering
\begin{overpic}[width=6.3cm,grid=false]{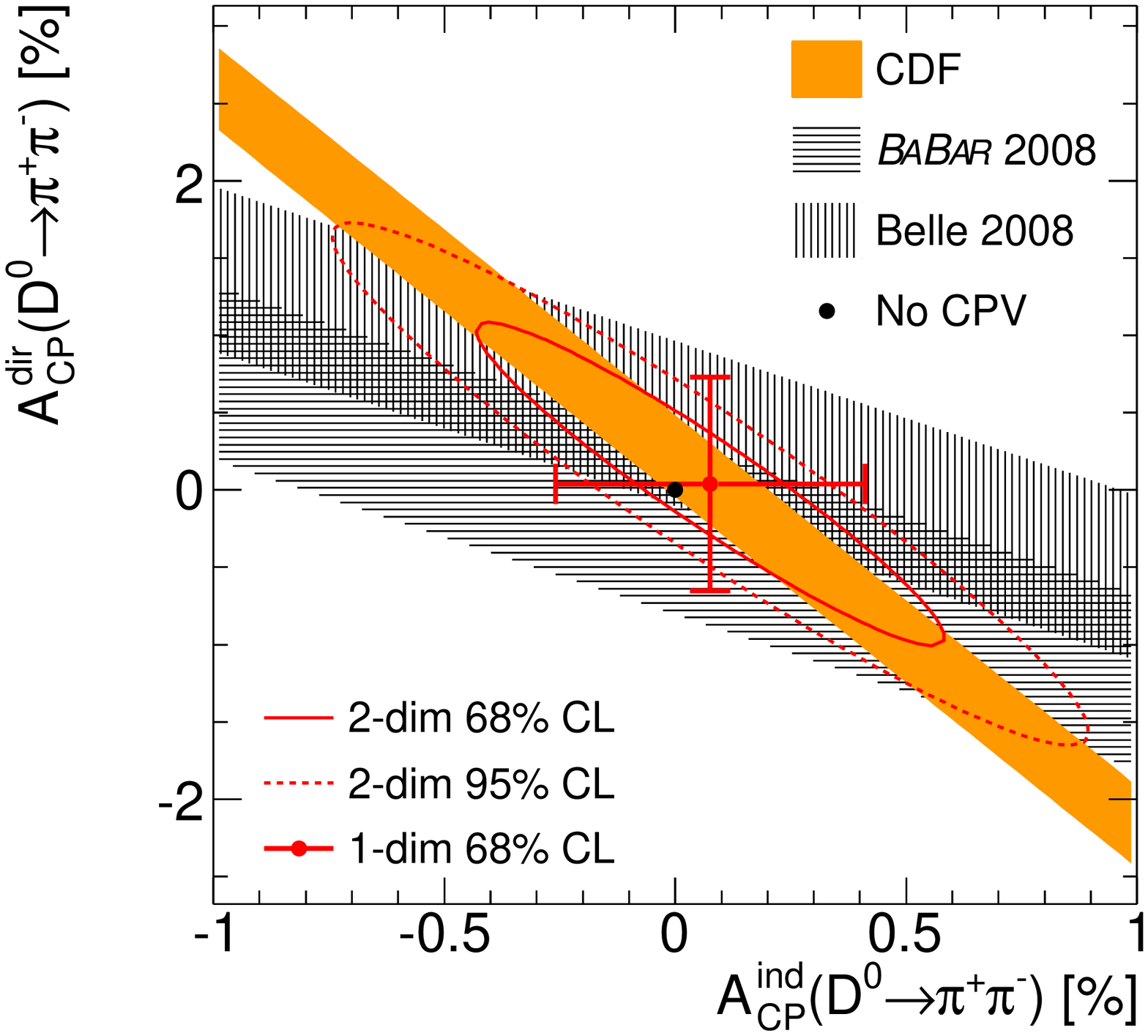}
\put(50,140){(a)} 
\end{overpic} \hfil
\begin{overpic}[width=6.3cm,grid=false]{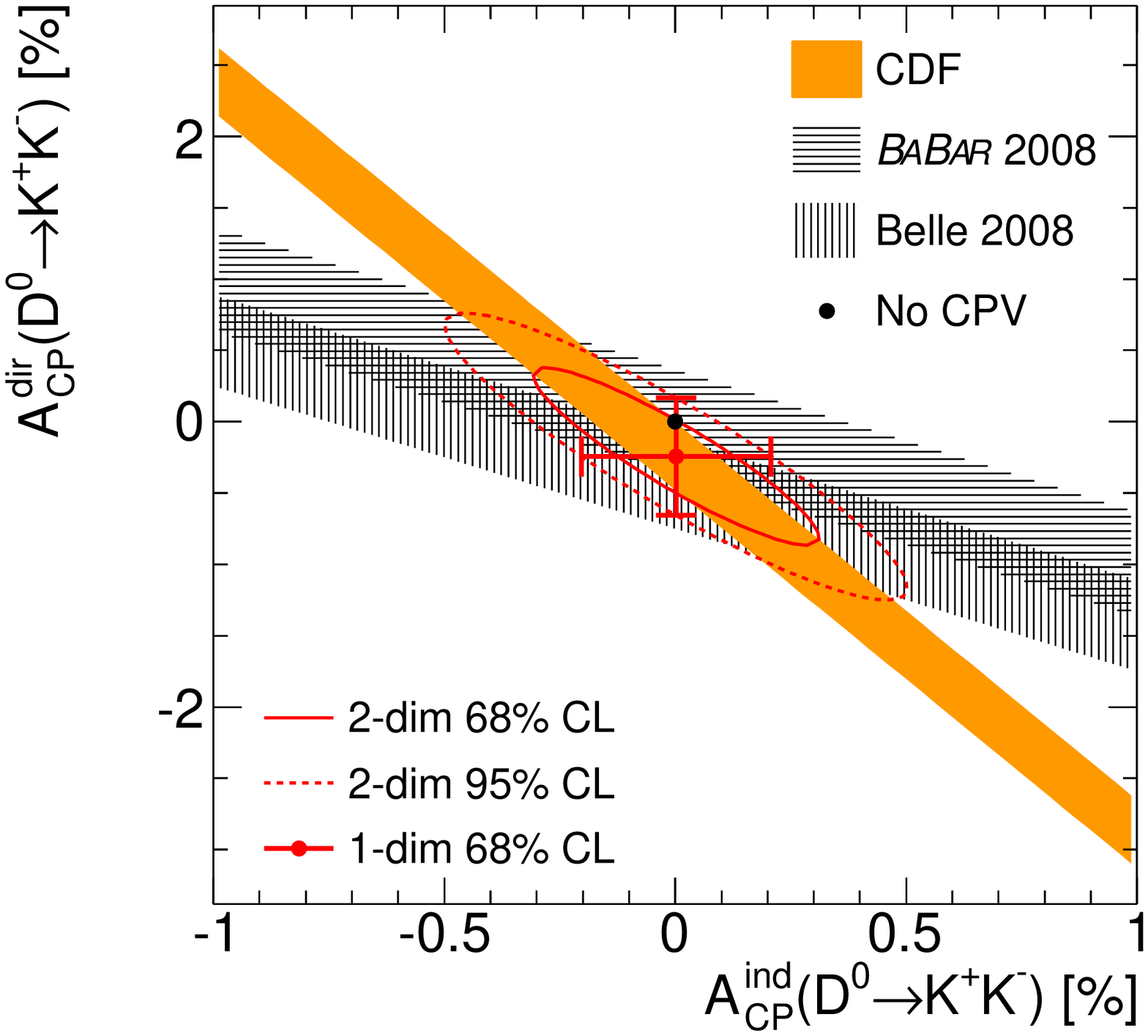}
\put(50,140){(b)}
\end{overpic}
\caption[xxx]{Comparison of the present results with Belle, \babar\ and CDF 
measurements of time-integrated \CP--violating asymmetry in (a) $D^0\to \pi^+\pi^-$  and (b) $D^0\to K^+K^-$  decays displayed 
in the $(\Acp^{\rm{ind}},\Acp^{\rm{dir}})$ plane.  The point with error bars denotes the central value of the combination of the three measurements 
with one-dimensional 68\% confidence level uncertainties. Figures are extracted from Ref.~\refcite{Aaltonen:2011se}.
Very recent measurement from Belle~\cite{d0hh_belle_2012} is not included in the average.}\label{fig:combination}
\end{figure*}
Each measurements defines a band in the  $(\Acp^{\rm{ind}},\Acp^{\rm{dir}})$  
plane with slope  $-\left<t\right>/\tau$ (Eq.\ (\ref{acp-time-integrated})).  CDF determines a mean decay time 
of $2.40\pm0.03$ and $2.65\pm0.03$~\cite{Aaltonen:2011se}, in units of $D^0$ lifetime, 
for \mbox{$D^0\to\pi^+\pi^-$} and \mbox{$D^0\to K^+K^-$} decays, respectively. 
%The uncertainty is the sum in 
%quadrature of statistical and systematic contributions. The small difference in the two samples is caused by the slightly 
%different kinematic distributions of the two decays, which impacts their trigger acceptance. 
The same holds for \babar\ and Belle measurements, with slope $-1$ \cite{Aubert:2007wf,Staric:2007dt}, due to unbiased acceptance in decay time. 
The most recent
% CDF and $B$-factories' 
results are shown in Fig.~\ref{fig:combination},  
which displays their relationship. The bands represent $\pm 1\sigma$ uncertainties and show 
that all measurements are compatible with \CP\ conservation (origin in the two-dimensional plane). The results of the three 
experiments can be combined assuming Gaussian uncertainties,  to construct a combined confidence regions in 
the $(\Acp^{\rm{ind}},\Acp^{\rm{dir}})$ plane, denoted with  $68\%$ and $95\%$ confidence level ellipses. The corresponding values for 
the asymmetries are $\Acp^{\rm{dir}}(D^0\to\pi^+\pi^-) = (0.04 \pm 0.69)\%$,   $\Acp^{\rm{ind}}(D^0\to\pi^+\pi^-) = (0.08 \pm 0.34)\%$,  
$\Acp^{\rm{dir}}(D^0\to K^+K^-) = (-0.24 \pm 0.41)\%$, and  $\Acp^{\rm{ind}}(D^0\to K^+K^-) = (0.00\pm 0.20)\%$, 
 in which the uncertainties represent one-dimensional 68\% confidence level intervals.

\section{Measurement of $\Delta\Acp$}
\label{ssec_delta_acp}

At LHC  charm and anticharm mesons  are produced by \CP--conserving 
strong interactions, through $pp$ initial state, which has a net charge of +2$e$, which is not \CP\ symmetric.
This produces a small net effect different from zero in the asymmetry production $A_p$, difficult to measure or cancel out in the measurement 
of the individual CP asymmetries. The production asymmetry cannot be ``easily''  disentangled from the intrinsic CP asymmetry, and 
so far LHCb did not provided any individual measurement of the time--integrated CP asymmetry in the singly-Cabibbo-suppressed decay modes.  
However this term cancels out in the measurement of the difference 
$$\Delta\Acp \equiv \Acp(D^0\to K^+K^-) - \Acp(D^0\to \pi^+\pi^-),$$
which could be maximally sensitive to \CP\ violation 
since the individual asymmetries are expected to have opposite sign, if the invariance  under the interchange 
of $d$ with $s$ quark is even approximately valid~\cite{Grossman:2006jg}.
%
%As mentioned in the pre the measured time-integrated asymmetry 
%depends on the time-acceptance of the experiment performing the measurement. 
%It can be written as
%\begin{equation}
%A_{\CP}(f) = \adirCP(f) \, + \, \frac {\langle t \rangle}{\tau} \aindCP, \label{eq:acpphysics}
%\end{equation}
%where $\langle t \rangle$ is the average decay time in the reconstructed sample.
From Eq.~\ref{acp-time-integrated}  the difference $\Delta\Acp$ can be written as 
%\begin{eqnarray}
%\Delta A_{\CP} & \equiv & A_{\CP}(D^0\to K^-K^+) \, - \, A_{\CP}(D^0\to \pi^-\pi^+) \label{eq:acpfinal} \\
%&  = & \left[ \Acp^{\rm dir}(K^-K^+) \,-\, \Acp^{\rm dir}(\pi^-\pi^+) \right] \, + \, \frac {\Delta \langle t \rangle}{\tau} \Acp^{\rm ind}.\nonumber 
%\end{eqnarray}
\begin{equation}
\Delta A_{\CP}  =  \left[ \Acp^{\rm dir}(K^-K^+) \,-\, \Acp^{\rm dir}(\pi^-\pi^+) \right] \, + \, \frac {\Delta \langle t \rangle}{\tau} \Acp^{\rm ind}
= \Delta A^{\rm dir}_{CP}   +  \frac {\Delta \langle t \rangle}{\tau} \Acp^{\rm ind},\label{eq:acpfinal}
\end{equation}
and in the limit that  $\Delta \langle t \rangle$ vanishes,
$\Delta A_{\CP}$ is equal to the difference in the direct \CP\ asymmetry
between the two decays $\Delta A^{\rm dir}_{\CP}$.
%However, if the time-acceptance is different for the $\Km\Kp$ and $\pim\pip$ final
%states, a contribution from indirect \CP violation remains.
%
\begin{table}[h]
\tbl{Summary of recent experimental measurements of $\Delta \Acp$ 
in two--body singly--Cabibbo--suppressed decays of $D^0$ mesons. The \babar\ measurement 
has been calculated assuming the two individual asymmetries uncorrelated.} %\cite{Aubert:2007if,:2008rx,Acosta:2004ts}.} 
{\begin{tabular}{lcc}
\toprule  
Experiment      & $\Acp(D^0\to K^+K^-) -\Acp(D^0\to \pi^+\pi^-)\ (\%)$ & significance \\
\colrule 
\babar\ 2008 \cite{Aubert:2007if}      & $-0.24\pm0.62 \pm0.26$  &  0.4$\sigma$\\
\belle\ 2008 \cite{Staric:2008rx}        & $-0.86\pm0.60 \pm0.07$ & 1.4$\sigma$\\
LHCb 2012 \cite{Aaij:2011in}              & $-0.82\pm0.21\pm0.11 $  & 3.5$\sigma$\\
CDF 2012 \cite{Aaltonen:2012qw}           & $-0.62\pm0.21\pm0.10$   & 2.7$\sigma$\\  %%%% correction proofs 
\belle\ 2012 \cite{d0hh_belle_2012}   & $-0.87\pm0.41 \pm0.06$ & 2.1$\sigma$\\
\botrule
\end{tabular} \label{today_delta}}
\end{table}
Tab.~\ref{today_delta} reports the most recent measurements of $\Delta A_{\CP}$ with the relative uncertainties, 
where the deviation from zero is calculated  by adding in quadrature the statistical and systematic 
uncertainty, assumed to be independent and Gaussian-distributed.
All measurements are consistent. In particular LHCb result strongly indicate, for the first time,  
the presence of CP violation in the charm sector, since $\Delta A_{\CP}$ deviates from zero by and 3.5$\sigma$,
confirmed by CDF results 2.7$\sigma$ from zero. The two results
have comparable accuracy and  less than 1$\sigma$ different in central value. The combined results of the two experiments provide 
substantial evidence for CP violation in the charm sector with a size larger
than most predictions \cite{LeYaouanc:1992iq,Buccella:1994nf}, possibly suggestive of the presence of non-SM dynamics.
Figure~\ref{hfag_cpv_plot}  shows the $\Delta A_{CP}$ measurements as a function 
of $\Acp^{\rm ind}$.

\section{Measurement of $A_{\Gamma}$}
\label{a_gamma}

The singly-Cabibbo-suppressed \Dhh\ decays probes  CP violation effects also through the observable 
$A_{\Gamma}$, given by the asymmetry of effective lifetimes as
\begin{equation}
A_{\Gamma} \equiv \frac{\tau(\overline{D}^0\to h^+ h^-)-\tau(D^0\to h^+h^-)}{\tau(\overline{D}^0\to h^+h^-)+\tau(D^0\to h^+h^-)}.
\label{eqn:agamma}
\end{equation}
where effective lifetime refers to the value measured using a single exponential model. 
Given the experimental constrains $x,y\ll 1$ and assuming $|\overline{A}_f/A_{f}| \approx 1$
one can write~\cite{Grossman:2006jg}:
%$A^{\rm dir}_{CP} \ll A^{\rm ind}_{CP}$ one can write~\cite{Grossman:2006jg}:
\begin{equation}
%A_{\Gamma} = -A^{\rm ind}_{CP}=\frac{1}{2} A^{\rm mix}_{CP} y \cos\phi - x\sin\phi.
A_{\Gamma} \approx -A^{\rm ind}_{CP}.
\end{equation}
%where 
%\begin{equation}
%A^{\rm mix}_{CP} = - \eta_{CP} \frac{y}{2} \left(       \left|\frac{q}{p}\right|  + \left|\frac{p}{q}\right|     \right) \sin\phi.
%\end{equation}
The measurement of $A_{\Gamma}$ is, therefore, described in most literature as a determination of indirect \CP\ violation.
However, because of the very recent measurements of $\Delta A_{CP}$,  the direct \CP\
 violation, which seems to be at the level of $10^{-2}$ cannot be neglected in the calculations. It 
 can have a contribution to $A_{\Gamma}$ at the level of $10^{-4}$.
 $A_{\Gamma}$ can be then expressed~\cite{Gersabeck:2011xj} as
%Using the same expansion as for $y_{CP}$ leads to
%\begin{eqnarray}
%A_{\Gamma} & \approx & \bigg[\frac{1}{2}(A^{\rm ind}_{CP}+A^{\rm dir}_{CP})y\cos\phi-x\sin\phi \bigg]\frac{1}{1+y_{CP}}\nonumber\\
%& \approx & \frac{1}{2}(A^{\rm ind}_{CP}+A^{\rm dir}_{CP})y\cos\phi-x\sin\phi.
%\label{eqn:agammatheory}
%\end{eqnarray}
\begin{equation}
%A_{\Gamma}=\frac{1}{2}(A^{\rm mix}_{CP}+A^{\rm dir}_{CP})y\cos\phi-x\sin\phi = - A^{\rm ind}_{CP} + A^{\rm dir}_{CP}\frac{1}{2}y\cos\phi.
A_{\Gamma} \approx - A^{\rm ind}_{CP} - A^{\rm dir}_{CP}\eta_{CP}y\cos\phi \approx - A^{\rm ind}_{CP} - A^{\rm dir}_{CP}y_{CP}. \label{Agamma_full}
\end{equation}
where $y_{CP}$ is the deviation from unity of the ratio of effective lifetimes in the decay modes $D^0\to h^+h^-$ and $D^0\to K^-\pi^+$
\begin{equation}
y_{\CP} \equiv \frac{\tau(D^0\to K^+ K^-)}{\tau(D^0\to K^- \pi^+)}-1.
\label{eqn:ycp}
\end{equation}
In the limit of no \CP\ violation $y_{CP}$ is equal to $y$ and hence becomes a pure mixing parameter.
%However, once precise measurements of $y$ and $y_{CP}$ are made, any difference between $y$ and $y_{CP}$ would be a sign of \CP\ violation.
%
The most recent measurement for $A_\Gamma$ and $y_{CP}$ are reported in Tab.~\ref{today_a_gamma}. The latest from 
\babar\ and \belle\  were presented very recently in 2012, confirming the presence of charm mixing, respectively at 3.3$\sigma$ and $4.8\sigma$,
and a value for $A_{\Gamma}$ consistent with zero.
\begin{table}[h]
\tbl{Summary of recent experimental measurements of $y_{CP}$ and $A_{\Gamma}$ 
in two body singly--Cabibbo--suppressed decays of the $D^0$ mesons. }
{\begin{tabular}{lcc}
\toprule  
Experiment                                             & $y_{CP}$ [\%] & $A_{\Gamma}$ [\%] \\
\colrule 
\babar\ 2007 \cite{Aubert:2007en}         & $1.24\pm 0.39 \pm 0.13$  &  $-0.26\pm 0.36 \pm0.08$\\
\belle\  2007 \cite{Staric:2007dt}           & $1.31\pm 0.32 \pm 0.25$  &  $+0.01\pm0.30 \pm0.15$ \\
LHCb 2012 \cite{Aaij:2011ad}                 & $0.55\pm0.63\pm0.41 $   &  $-0.59\pm0.59 \pm0.21$ \\
\babar\ 2012 \cite{Lees:2012qh}       & $0.72\pm0.18\pm0.12 $   &  $+0.09\pm0.26 \pm0.06$ \\
\belle\ 2012 \cite{staric_charm}             & $1.11\pm0.22\pm0.11 $   &  $-0.03\pm0.20 \pm0.08$ \\
\botrule
\end{tabular} \label{today_a_gamma}}
\end{table}

As for the measurement of the time-integrated CP asymmetry, it is
necessary to  reconstruct the  $D^{\ast +}\to D^0\pi^+_s$ decays with a characteristic
slow pion $\pi_s$, and $D^0\to K^+K^-$, $K^-\pi^+$, and $\pi^+\pi^-$.
To select pion and kaon candidates, standard particle identification criteria are imposed.
$D^0$ daughter tracks are refitted to  a common vertex, and the $D^0$ production vertex is
found by constraining its momentum vector and the $\pi_s$ track to originate
from the $e^+e^-$ interaction region. %; confidence levels exceeding $10^{-3}$
%are required for both fits.  A $D^\ast$ momentum greater than 
%$2.5\,\gev/c$ (in the CM) is required to reject $D$-mesons
%produced in $B$-meson decays and to suppress combinatorial background.
The proper  decay time of the
$D^0$ candidate is then calculated from the projection of the vector
joining the two vertices, $\vec{L}$, onto the $D^0$ momentum vector, 
$t=m_{D^0}\vec{L}\cdot\vec{p}/p^2$, where
$m_{D^0}$ is the nominal $D^0$ mass. The decay time uncertainty 
$\sigma_t$ is evaluated event-by-event from the covariance
matrices of the production and decay vertices.
%
%Candidate $D^0$ mesons are selected using two kinematic observables:
%the invariant mass of the $D^0$ decay products, $M$, and the energy
%released in the $D^{\ast +}$ decay,
%$q=(M_{D^\ast}-M-m_\pi)c^2$. $M_{D^\ast}$ is the invariant mass of
%the $D^0 \pi_s$ combination and $m_\pi$ is the $\pi^+$ mass.
\begin{figure*}[t]
\begin{center}
\includegraphics[width=8.57cm]{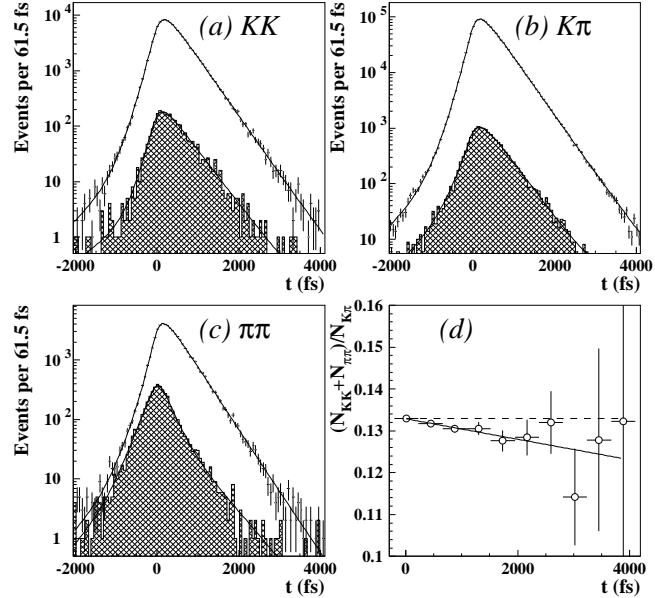}
\caption[xxx]{Results of the simultaneous fit to decay time distributions
of (a) $D^0\to K^+K^-$, (b) $D^0\to K^-\pi^+$ and 
(c) $D^0\to \pi^+\pi^-$ decays from Ref.~\refcite{Staric:2007dt}. The
cross-hatched area represents background contributions, the shape of
which was fitted using mass sideband events.
(d) Ratio of decay time distributions between $D^0\to K^+K^-,
\pi^+\pi^-$ and $D^0\to K^-\pi^+$ decays. The solid line is a fit to the data
points.}
 \label{belle_Agamma}
\end{center}
\end{figure*} 
Figure~\ref{belle_Agamma} reports the results of the
simultaneous fit to to decay time distributions of  $D^0\to K^+K^-$, $D^0\to K^-\pi^+$ and 
$D^0\to \pi^+\pi^-$ decays at Belle~\cite{Staric:2007dt}. Similar
distributions are obtained at \babar\ and LHCb.  For all experiments
the main challenges are the extraction of the time resolution function
and the acceptance variations with the decay time. In particular the
last is crucial in LHCb where a lifetime-biasing selection is applied.    

\begin{figure*}[t]
\centering
%\begin{overpic}[width=6.3cm,grid=false]{./fig/deltaAcp_AGamma_combination.eps}
%\begin{overpic}[width=8cm,grid=false]{./fig/deltaACP_AGamma_fit_06mar2012.eps}
\begin{overpic}[width=8cm,grid=false]{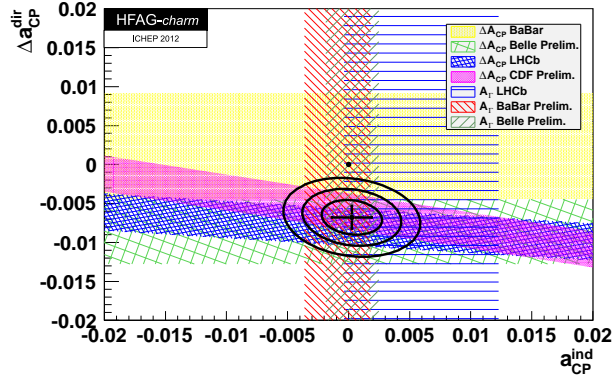}
%\put(23,34){(c)}
\end{overpic}
\caption[xxx]{Representation of the current knowledge on \CP\ violation in \Dhh\ decays in the plane $(\Acp^{\rm ind}, \Delta\Acp^{\rm dir})$.
The combination of all results assumes Gaussian, fully uncorrelated
uncertainties. See Ref.~\refcite{hfag} for details of combination and references therein.
%Refs.~\refcite{Aubert:2007if,d0hh_belle_2012,Aaltonen:2011se,Aaij:2011in,Aubert:2007en,Staric:2007dt,Aaij:2011ad,Lees:2012qh,staric_charm}.
%for the individual measurements. Measurement of $A_{\Gamma}$ from Ref.~\refcite{Lees:2012qh} and Ref.~\refcite{staric_charm}
%are very recent and are not reported.
} \label{hfag_cpv_plot}
\end{figure*}
Figure~\ref{hfag_cpv_plot}  shows the current knowledge on \CP\ violation in \Dhh\ decays in the plane $(\Acp^{\rm ind}, \Delta\Acp^{\rm dir})$, 
where all measurements of $\Delta A_{CP}$ and $A_{\Gamma}$ are reported.
%~\footnote{The \babar\ and \belle\ measurements 
%of $A_{\Gamma}$ reported are from Ref. \refcite{Aubert:2007en} and Ref.~\refcite{Staric:2007dt}. The latest measurement of 2012 are not 
%in the figure. }. 
A $\chi^2$ fit is performed to combine all measurements reported in Tab.\ref{today_delta} and \ref{today_a_gamma}. 
Statistical and systematic uncertainties are added in quadrature 
when calculating the $\chi^2$. The current world average value $y_{CP} = (1.064\pm 0.209)\%$~\cite{hfag} is used since it appears 
in the expression for $A_{\Gamma}$ of Eq.~\ref{Agamma_full}.  Moreover using the same approximation of Eq.~\ref{Agamma_full} 
the difference of time-integrated CP asymmetries of Eq.~\ref{eq:acpfinal} must be written as 
\begin{equation}
\Delta \Acp \approx \Delta \Acp^{\rm dir} \left(1 + y_{CP} \frac{\overline{\langle t \rangle}}{\tau}\right) 
+ \frac {\Delta \langle t \rangle}{\tau} \Acp^{\rm ind}
\end{equation}
where
$\Delta\langle t \rangle $ is the difference between the averaged quantity $\langle t \rangle$ for the $KK$ and $\pi\pi$ final state and  $\overline{\langle t \rangle}$ is their average.
%and $\overline{\langle t \rangle}$ denotes the average for quantity $\langle t \rangle$. 
%
%\begin{equation}
%\Delta A_{\CP}  =  \left[ \Acp^{\rm dir}(K^-K^+) \,-\, \Acp^{\rm dir}(\pi^-\pi^+) \right] \, + \, \frac {\Delta \langle t \rangle}{\tau} \Acp^{\rm ind},\label{eq:aacpfinal}
%\end{equation}
%and in the limit that  $\Delta \langle t \rangle$ vanishes,
%$\Delta A_{\CP}$ is equal to the difference in the direct \CP\ asymmetry
%between the two decays.
%
% and the measurements listed in the table below.  
%The combination plot shows the measurements listed in the table above for ΔACP and AΓ, where 
The bands represent $\pm1\sigma$ intervals, the point of no CP violation $(0,0)$ is shown as a filled circle, 
and two-dimensional 68\% C.L., 95\% C.L., and 99.7\% C.L. regions are plotted as ellipses with the best fit 
value as a cross indicating the one-dimensional uncertainties in their center.
From the fit, the change in $\chi^2$ from the minimum value for the no-CPV point (0,0) is 21.7; 
this corresponds to a C.L. of $2 \times 10^{-5}$ for two degrees of freedom. Thus the data is consistent 
with no CP violation at 0.002\% C.L. The central values and $\pm 1\sigma$ errors for the individual parameters are:
\begin{eqnarray}
 \Acp^{\rm ind}  & = & (-0.027 \pm 0.163 )\%   \\
\Delta\Acp^{\rm dir} & = & (-0.678 \pm 0.147 )\%
\end{eqnarray}
%The latest measurement\cite{Lees:2012qh,staric_charm} of $A_\Gamma$ and $y_{CP}$ from \babar\ and \belle\ are not included in this fit. 
Details can be found in Ref.~\refcite{hfag}.

%
%
%
%Previous measurements of $y_{CP}$ have been performed by \babar and \belle.
%The results are $y_{CP}=(11.6\pm2.2\pm1.8)\times 10^{-3}$~\cite{Aubert:2009ck} for \babar and $y_{CP}=(13.1\pm3.2\pm2.5)\times 10^{-3}$~\cite{Staric:2007dt} 
%for \belle. They are consistent with the world average of $y=(7.5\pm1.2)\times 10^{-3}$~\cite{hfag}.
%
%The study of the lifetime asymmetry of $D^0$ and $\overline{D}^0$ mesons decaying into the singly Cabibbo-suppressed 
%final state $K^+K^-$ can reveal indirect \CP\ violation in the charm sector.

%\section{CP violation in multibody D decays}

%dfkjslfjlsjflsjls

\section{Conclusions}

Recent charm physics measurements reached for the first time an interesting precision 
after many years of the discovery of $c$ quark. The size of the available data samples of charmed mesons decays 
allows a first exploration of the the Standard Model predictions in this territory, in fact the 
first evidence for \CP\ violation  %in the charm sector 
has already  opened a privileged door for probing effects of New Physics. %beyond the standard model.
The observation of the \CP\ violation in the charm sector is a near term goal,  achievable in a short time scale, if current hints will be confirmed. 
Instead, the real long term challenge will be the interpretation of the observed effects, where the relatively small
charm quark mass and the large cancellations in this system, makes it very hard.
In particular more precise determinations of the individual asymmetries in $D^0 \to \pi^+\pi^-$ and $D^0 \to K^+K^−$ decays and extension 
of the precise experimental exploration to other charm decays may help in understanding whether the 
observed effect can be attributed to significant hadronic corrections to the SM weak amplitudes or to new, non-SM sources of \CP\ violation.
Therefore precise measurements of both time-dependent and time-integrated asymmetries are necessary to reveal the nature of \CP\
violating effects in the $D^0$ system.

Since B factories, CLEO-c and CDF are analysing their final datasets, while LHCb and BESIII are currently taking data,  new results are expected to come soon. 
However to deeply explore the very interesting territory of charm CP violation, we will need the next generation experiments, the LHCb upgrade and the 
the future $e^+e^−$ collider experiments.

% \appendix

% \section{Appendices}

% Appendices should be used only when absolutely necessary. They
% should come before Acknowledgments. If there is more than one
% appendix, number them alphabetically. Number displayed equations
% occurring in the Appendix in this way, e.g.~(\ref{appeqn}), (A.2),
% etc.
% \begin{equation}
% \mu(n, t) = \frac{\sum^\infty_{i=1} 1(d_i < t, N(d_i) 
% = n)}{\int^t_{\sigma=0} 1(N(\sigma) = n)d\sigma}\,.
% \label{appeqn}
% \end{equation}

% \section*{Acknowledgments}

% This section should come before the References. Dedications and funding 
% information may also be included here.

% \section*{References}

% References are to be listed in the order cited in the text in Arabic
% numerals.  They can be typed in superscripts after punctuation marks,
% e.g.~``$\ldots$ in the statement.\cite{Toimela}'' or used directly,
% e.g.~``see Ref.~\refcite{Bohr} for examples.''  Please list using the
% style shown in the following examples.  For journal names, use the
% standard abbreviations.  Typeset references in 9 pt Times Roman. 
% Each reference number should consist of one reference only.


\begin{thebibliography}{0}


\bibitem{cpv_kmeson} J.H. Christenson \etal, \textit{Phys. Rev. Lett.} \textbf{13}, 138 (1964).

\bibitem{Aubert:2001nu} B. Aubert \etal, (\babar\ Collaboration), \textit{Phys. Rev. Lett.} \textbf{87}, 091801 (2001).

\bibitem{Abe:2001xe} K. Abe \etal, (\belle\ Collaboration), \textit{Phys. Rev. Lett.} \textbf{87}, 091802 (2001).

\bibitem{Burkhardt:1988yh} H. Burkhardt \etal, (NA31 Collaboration), \textit{Phys. Lett.} \textbf{B206}, 169 (1988).

\bibitem{Fanti:1999nm} V. Fanti \etal, (NA48 Collaboration), \textit{Phys. Lett.} \textbf{B465}, 335 (1999).

\bibitem{ktev} A. Alavi-Harati \etal, (KTeV Collaboration), \textit{Phys. Rev. Lett.} \textbf{83}, 22 (1999).

\bibitem{Aubert:2004qm} B. Aubert \etal, (\babar\ Collaboration), \textit{Phys. Rev. Lett.} \textbf{93}, 131801 (2004).

\bibitem{belle_acp}  S.-W. Lin \etal, (Belle Collaboration), \textit{Nature} \textbf{452}, 332 (2008)
% K. Abe \etal, (Belle Collaboration), arXiv:hep-ex/0507045.

\bibitem{Aubert:2005ce} B. Aubert \etal, (\babar\ Collaboration), \textit{Phys. Rev.} \textbf{D72}, 072003 (2005).

\bibitem{Garmash:2005rv} A. Garmash \etal, (Belle Collaboration), \textit{Phys. Rev. Lett.} \textbf{96}, 251803 (2006).

\bibitem{km_mechanism} M. Kobayashi and T. Maskawa, \textit{Prog. Theor. Phys.} \textbf{49}, 652 (1973).

\bibitem{Nir:1993mx} Y. Nir and N. Seiberg, \textit{Phys. Lett.} \textbf{B309}, 337 (1993). 

\bibitem{Ciuchini:2007cw} M. Ciuchini \etal, \textit{Phys. Lett.} \textbf{B655}, 162 (2007).

\bibitem{Aubert:2007wf} B. Aubert \etal, (\babar\ Collaboration), \textit{Phys. Rev. Lett.} \textbf{98}, 211802 (2007).

\bibitem{Staric:2007dt} M. Staric \etal, (Belle Collaboration), \textit{Phys. Rev. Lett.} \textbf{98}, 211803 (2007).

\bibitem{Aaltonen:2007uc} T. Aaltonen \etal, (CDF Collaboration), \textit{Phys. Rev. Lett.} \textbf{100}, 121802 (2008).

%\bibitem{hfag}D. Asner \etal, arXiv:1010.1589 and online update at \texttt{http://www.slac.stanford.edu/xorg/hfag}.

\bibitem{hfag} Y. Amhis \etal, arXiv:1207.1158 [hep-ex] and online update at \texttt{http://www.slac.stanford.edu/xorg/hfag}.

%Heavy Flavor Averaging Group, , “Averages of b-hadron, c-hadron, and τ -lepton properties as of early 2012,” arXiv:1207.1158 [hep-ex]. 
%Online update at http://www.slac.stanford.edu/xorg/hfag.


\bibitem{Petrov:2006nc} A. A. Petrov, \textit{Int. J. Mod. Phys. A} \textbf{21}, 5686 (2006).

\bibitem{Golowich:2007ka} E. Golowich, J. Hewett, S. Pakvasa, and A. A. Petrov, \textit{Phys. Rev.} \textbf{D76}, 095009 (2007).

\bibitem{Nir:1992uv} Y. Nir, Conf.Proc. \textbf{C9207131}, 81 (1992).

\bibitem{Bianco:2003vb}
  S.~Bianco, F.~L.~Fabbri, D.~Benson, and I.~I.~Bigi,  \textit{Riv.\ Nuovo Cim.}  {\bf 26N7}, 1 (2003).

\bibitem{Grossman:2006jg}
Y.~Grossman, A.~L.~Kagan, and Y.~Nir, \textit{Phys.\ Rev.} {\bf D75}, 036008 (2007).

\bibitem{Bigi:1986dp}
I.~I.~Bigi and A.~I.~Sanda, \textit{Phys.\ Lett.\ B}   {\bf 171}, 320 (1986).

%%%% BABAR ACP(D0->hh)
\bibitem{Aubert:2007if} B. Aubert \etal, (\babar\ Collaboration), \textit{Phys. Rev. Lett.} \textbf{100}, 061803 (2008).

%%%% Belle ACP(D0->hh)
\bibitem{Staric:2008rx} M. Staric \etal, (\belle\ Collaboration), \textit{Phys. Lett.} \textbf{B670}, 190 (2008).

%%%% CDF ACP(D0->hh)
\bibitem{Aaltonen:2011se} T. Aaltonen \etal, (CDF Collaboration), \textit{Phys. Rev. D} \textbf{85}, 012009 (2012).

%%%% Belle ACP(D0->hh) 2012
%\bibitem{d0hh_belle_2012} Byeong Rok Ko \etal, (Belle Collaboration), Direct CP Violation in charm at Belle, 
%talk at ICHEP 2012, Melbourne, Australia.
\bibitem{d0hh_belle_2012} B. R. Ko \etal, (Belle Collaboration), Direct CP Violation in charm at Belle, 
talk at ICHEP 2012, Melbourne, Australia.

%%%% CDF Delta_ACP(D0->hh)
%\bibitem{cdfnote_10784} T. Aaltonen \etal, (CDF Collaboration), \textit{CDF Public Note} 10784 (2012).
\bibitem{Aaltonen:2012qw} T. Aaltonen \etal, (CDF Collaboration), \textit{Phys. Rev. Lett.} \textbf{109}, 111801 (2012).

%%%% LHCb Delta_ACP(D0->hh)
\bibitem{Aaij:2011in} R. Aaij \etal, (LHCb Collaboration), \textit{Phys. Rev. Lett.} \textbf{108}, 111602 (2012).

\bibitem{afb}  F.A. Berends, K.J.F. Gaemers, R. Gastmans, \textit{Nucl. Phys. B}{\bf 63}, 381 (1973). 
\bibitem{afb1}  R.W. Brown, K.O. Mikaelian, V.K. Cung, E.A. Paschos, \textit{Phys. Lett. B}{\bf 43}, 403 (1973).
\bibitem{afb2}  R.J. Cashmore, C.M. Hawkes, B.W. Lynn, R.G. Stuart, \textit{Z. Phys. C}{\bf 30}, 125 (1986).  

\bibitem{LeYaouanc:1992iq} A. Le Yaouanc, L. Oliver and  J.C. Raynal,
\textit{Phys.\ Lett.\ }{\bf B292}, 353 (1992).

\bibitem{Buccella:1994nf} F. Buccella \etal, 
\textit{Phys.\ Rev.\ D }{\bf 51}, 3478 (1995).

% A_Gamma

%% vedi sopra
%\bibitem{Grossman:2006jg} Y. Grossman, A. Kagan and Y. Nir, \textit{Phys. Rev. D} \textbf{75}, 036008 (2006).
%New physics and CP violation in singly Cabibbo  suppressed D decays
% hep-ph/0609178 

\bibitem{Gersabeck:2011xj} M. Gersabeck \etal, \textit{J. Phys. G}\textbf{39}, 045005 (2012).
%On the interplay of direct and indirect CP violation in the charm sector
%  eprint         = "1111.6515",
%   archivePrefix  = "arXiv",
%   primaryClass   = "hep-ex",

\bibitem{Aubert:2007en} B. Aubert \etal, (\babar\ Collaboration), \textit{Phys. Rev. D} \textbf{78}, 011105 (2008).

\bibitem{Aaij:2011ad} R. Aaij \etal, (LHCb Collaboration), \textit{JHEP} \textbf{1204}, 129 (2012).


%\bibitem{casarosa_charm} G. Casarosa,  Measurement of $D^0-\overline{D}^0$ mixing and CP violation at Babar, talk at CHARM 2012, 
%Honolulu, Hawai'i, USA.
\bibitem{Lees:2012qh} Lees, J.P. \etal,  (\babar\ Collaboration), arXiv:1209.3896 [hep-ex].
%Measurement of $D^0-\overline{D}^0$ mixing and CP violation at Babar, talk at CHARM 2012, 
%Honolulu, Hawai'i, USA.

% @article{Lees:2012qh,
%       author         = "Lees, J.P. and others",
%       title          = "{Measurement of $D^0-\bar{D}^0$ Mixing and CP Violation
%                         in Two-Body $D^0$ Decays}",
%       collaboration  = "BABAR Collaboration",
%       year           = "2012",
%       eprint         = "1209.3896",
%       archivePrefix  = "arXiv",
%       primaryClass   = "hep-ex",
%       SLACcitation   = "%%CITATION = ARXIV:1209.3896;%%",
% }

\bibitem{staric_charm} M. Staric,  New Belle reults on $D^0-\overline{D}^0$, talk at CHARM 2012, 
Honolulu, Hawai'i, USA.


%%% #########


% \bibitem{Marnelius} R. Marnelius, {\it Acta Phys. Pol. B} {\bf 13},  
% 669 (1982).

% \bibitem{Bjorken} J. D. Bjorken, in {\it Lecture Notes on Current-Induced 
% Reactions}, eds.~J. Komer {\it et al.} (Springer, 1975).

% \bibitem{Bohr} A. Bohr and B. R. Mottelson, {\it Nuclear Structure} 
% (Benjamin, 1969), Vol.~1, pp.~100--102.

% \bibitem{Webb} R. C. Webb, Ph.D. thesis, Princeton University, 1972.

% \bibitem{Toimela} T. Toimela, Helsinki Research Institute for 
% Theoretical Physics, Report No. HU-TFT-82-37, 1982 (unpublished).

\end{thebibliography}
\end{document}